\input phyzzx
\input maggieref
\newcount\mongocount
\mongocount=1
\def\Figure#1#2#3{
      \vbox to #3in{\hsize=#2in
        \vfil
         \includegraphics{#1}
    }
}
\def\figcap#1#2{
\vtop{\tenpoint\singlespace
\hsize=#1in\smallskip\noindent Figure\ \ \the\mongocount.\ \  #2
\global\advance\mongocount by 1\bigskip}}
\def\mongofigure#1#2#3#4#5{\centerline{\Figure{#1}{#2}{#3}
\figcap{#4}{#5}}}

\hoffset=0.375in
\overfullrule=0pt

\def\exp{{\rm exp}}
\def\eff{{\rm eff}}
\def\max{{\rm max}}
\def\sech{{\rm sech}}

\def\min{{\rm min}}

\def\pc{{\rm pc}}

\twelvepoint
\font\bigfont=cmr17
\centerline{\bigfont M Dwarfs From Hubble Space Telescope Star Counts III:}
\smallskip
\centerline{\bigfont The Groth Strip}
\bigskip
\centerline{{\bf Andrew Gould}\foot{Alfred P.\ Sloan Foundation Fellow}}
\smallskip
\centerline{Dept of Astronomy, Ohio State University, Columbus, OH 43210}
\smallskip
\bigskip
\centerline{\bf John N.\ Bahcall}
\smallskip
\centerline{Institute For Advanced Study, Princeton, NJ}
\bigskip 
\centerline{\bf Chris Flynn}
\smallskip
\centerline{Tuorla Observatory, Piikki\"{o}, FIN-21500, Finland}
\bigskip
\centerline{e-mail: gould@payne.mps.ohio-state.edu, jnb@ias.edu,
cflynn@astro.utu.fi}

\centerline{\bf Abstract}
\singlespace
We analyze the disk M dwarfs found in 31 new fields observed with the Wide 
Field Camera (WFC2) on the {\it Hubble Space Telescope (HST)} together with
the sample previously analyzed from 22 WFC2 fields and 162 prerepair
Planetary Camera (PC1) fields.  The new observations, which include the 28 
high-latitude fields comprising the Large Area Multi-Color Survey 
(``Groth Strip''), increase the total sample to 337 stars, and more than
double the number of late M dwarfs ($M_V>13.5$) from 23 to 47.  
The mass function changes slope at $M\sim 0.6 M_\odot$, from a near-Salpeter
power-law index of $\alpha=-1.21$ to $\alpha=0.44$.  In both regimes the 
mass function at the Galactic plane is given by
$${d^3 N\over d\log M\, d\,M_V\, d V}
=8.1\times 10^{-2}\pc^{-3}\biggl({M\over 0.59\,M_\odot}\biggr)^{\alpha}$$
The correction for
secondaries in binaries 
changes the low-mass index from $\alpha=0.44$ to $\alpha\sim 0.1$.  
If the Salpeter slope 
continued to the hydrogen-burning limit, we would expect 500 stars in the
last four bins $(14.5<M_V<18.5)$, instead of the 25 actually detected.  The 
explanation of the observed microlensing rate towards the Galactic bulge
requires either a substantial population of bulge brown dwarfs or that the
disk and bulge mass functions are very different for stars with 
$M\lsim 0.5\,M_\odot$.


Subject Headings:  dark matter -- gravitational lensing -- stars: low mass,
luminosity function
\smallskip
\centerline{submitted to {\it The Astrophysical Journal}: November 20, 1996}
\centerline{Preprint: OSU-TA-31/96}

\endpage

\chapter{Introduction}

	We present new results from a search for M dwarfs found in
images taken using the {\it Hubble Space Telescope (HST)}.  The primary aim
of this program is to measure the faint end of the disk luminosity function
(LF), and thus to address four questions:  What is the contribution of M stars
to the disk mass?  Is the disk mass function (MF) rising at the last measured
point which would indicate the possible presence of brown dwarfs beyond the 
hydrogen-burning limit?  What is the vertical distribution of the disk?  What 
contribution do disk stars make to the observed microlensing events?

	In Paper I (Gould, Bahcall, \& Flynn 1996) we analyzed a total of
257 M dwarfs, 192 in 22 fields imaged with the repaired
Wide Field Camera (WFC2) with mean limiting mag $I=23.7$ and 65 stars in
162 fields imaged with the prerepair Planetary Camera (PC1) with mean limiting
mag $V=21.3$.  Our principal result was that the disk LF peaks at $M_V\sim 12$
and correspondingly that the disk MF (per unit log mass) peaks at 
$M\sim 0.5\,M_\odot$.

	Here we combine these previous results with 80 additional
M dwarfs found in 31 WFC2 fields of which 28 slightly overlapping fields
comprise the Large Area Multi-Color Survey (``Groth Strip'', $l=96^\circ$, 
$b=60^\circ$).  The Groth Strip is at substantially higher latitude than
the typical field analyzed in Paper I, and therefore the mean number of
stars per field in the new sample is substantially smaller (2.6 vs.\ 8.7).  
However, 
high-latitude fields are relatively more sensitive to intrinsically fainter
stars, so the additional 31 new fields more than double the number of late M 
dwarfs ($M_V>13.5$) in the sample from 23 to 47.  Since the determination of
the LF is mainly limited by small number statistics at the faint end, these
additional fields permit a significant improvement in the measurement.

	Our principal result is to confirm the break in the MF with our
best estimate now at $M\sim 0.6$, and to quantify this break
as a change in the power-law index from the near-Salpeter
value of $\alpha=-1.21$ to $\alpha=0.44$ ($\alpha=-1.2$ to $\alpha=0.1$ after
correction for binaries).

\chapter{Observations and Analysis}

	Table 1 gives the characteristics of the 31 new WFC2 fields as
well as those of the 22 fields analyzed in Paper I.  The fields are listed
in order of increasing Galactic latitude in each group.  In Paper I we 
described our method for
determinating the detection threshold $(I_\max)$ and the saturation threshold 
$(I_\min)$.  Table 1 also gives the area of each field in units of the
effective area of WFC2 ($4.4\,\rm arcmin^2$).  

\midinsert
\singlespace
$$\vbox{\halign{\hfil#\quad
&\hfil#\quad&\hfil#\quad&\hfil#
&\hfil#\quad&\hfil#&\hfil#\cr
\multispan{7}{\hfil TABLE 1 \hfil}\cr
\noalign{\medskip}
\multispan{7}{\hfil WFC2 Fields \hfil}\cr
\noalign{\smallskip}
\noalign{\hrule}
\noalign{\smallskip}
\noalign{\hrule}
\noalign{\smallskip}
 \hfil RA \hfil 
& \hfil Dec\hfil 
& \hfil $l$\hfil 
& \hfil $b$\hfil 
& \hfil $I_{\rm max}$\hfil 
& \hfil $I_{\rm min}$\hfil 
& \hfil $\Omega$\hfil\cr
\noalign{\smallskip}
\hfil (2000) \hfil
& \hfil (2000) \hfil
& & & & &
\hfil (WFC2)\hfil\cr
\noalign{\smallskip}
\noalign{\hrule}
\noalign{\smallskip}
\multispan{7}{\hfil New Observations \hfil}\cr
\noalign{\smallskip}
\noalign{\hrule}
\noalign{\smallskip}
17 14 14.88 & $+$50 15 30.0 & $ 77$ & $+ 36$ &   24.72 &   19.42 & 0.90\cr
15 58 49.18 & $+$42 05 26.3 & $ 67$ & $+ 49$ &   24.95 &   19.45 & 1.00\cr
12 36 49.40 & $+$62 12 58.0 & $126$ & $+ 55$ &   25.67 &   19.45 & 1.00\cr
14 16 31.98 & $+$52 15 51.9 & $ 96$ & $+ 60$ &   23.79 &   18.75 &25.98\cr
\noalign{\smallskip}
\noalign{\hrule}
\noalign{\smallskip}
\multispan{7}{\hfil Previous Observations \hfil}\cr
\noalign{\smallskip}
\noalign{\hrule}
\noalign{\smallskip}
21 51 17.91 & $+$28 59 53.9 & $ 82$ & $- 19$ &   23.46 &   18.09 & 1.00\cr
21 51 34.68 & $+$28 58 13.3 & $ 82$ & $- 19$ &   23.59 &   18.84 & 1.00\cr
04 24 55.56 & $+$17 04 47.8 & $179$ & $- 22$ &   22.83 &   17.65 & 1.00\cr
06 52 43.16 & $+$74 21 38.4 & $140$ & $+ 26$ &   23.98 &   19.45 & 0.67\cr
07 42 41.12 & $+$49 44 17.5 & $169$ & $+ 28$ &   23.98 &   19.45 & 1.00\cr
07 42 44.66 & $+$65 06 08.5 & $151$ & $+ 30$ &   23.98 &   19.45 & 1.00\cr
00 49 06.99 & $+$31 55 48.6 & $122$ & $- 31$ &   23.98 &   19.45 & 1.00\cr
14 42 16.61 & $-$17 10 58.7 & $337$ & $+ 38$ &   23.73 &   17.05 & 1.00\cr
16 01 22.24 & $+$05 23 37.2 & $ 16$ & $+ 40$ &   23.37 &   18.64 & 1.00\cr
00 29 05.46 & $+$13 08 07.4 & $115$ & $- 49$ &   24.07 &   19.45 & 1.00\cr
03 49 58.89 & $-$38 13 43.3 & $241$ & $- 51$ &   24.40 &   17.89 & 1.00\cr
14 13 11.78 & $-$03 07 57.0 & $339$ & $+ 54$ &   23.22 &   18.53 & 1.00\cr
15 19 41.20 & $+$23 52 05.4 & $ 36$ & $+ 57$ &   24.26 &   18.84 & 1.00\cr
12 55 41.55 & $-$05 50 56.9 & $305$ & $+ 57$ &   23.84 &   19.45 & 1.00\cr
01 44 10.61 & $+$02 17 51.2 & $148$ & $- 58$ &   23.74 &   19.45 & 1.00\cr
14 45 10.26 & $+$10 02 49.7 & $  6$ & $+ 58$ &   22.91 &   17.34 & 1.00\cr
02 56 22.03 & $-$33 22 25.3 & $234$ & $- 62$ &   23.98 &   19.45 & 1.00\cr
13 38 18.49 & $+$04 28 03.1 & $331$ & $+ 65$ &   23.40 &   17.50 & 1.00\cr
01 10 03.01 & $-$02 26 22.8 & $134$ & $- 65$ &   23.58 &   19.45 & 1.00\cr
01 09 59.79 & $-$02 27 23.7 & $134$ & $- 65$ &   23.83 &   19.45 & 0.67\cr
14 34 51.89 & $+$25 10 04.5 & $ 34$ & $+ 67$ &   23.92 &   18.64 & 1.00\cr
01 17 07.71 & $-$08 39 10.9 & $142$ & $- 71$ &   22.56 &   17.89 & 1.00\cr
\noalign{\smallskip}
\noalign{\hrule}
}}
$$

\endinsert

	For most of the fields, the method used to identify and measure point 
sources is the same as 
in Paper I and is illustrated by figure 2b of Flynn, Gould, \& Bahcall (1996).
However, four of the fields (the first three listed in Table 1, plus one of
the Groth Strip fields) are dithered and required a new method of analysis.
This method is described by Flynn et al.\ (1996) and is illustrated by their
figure 2a.

The transformations from instrumental to standard Johnson-Cousins
mags are given by equation (2.1) of Bahcall et al.\ (1994).  As in Paper I
we adopt the color-mag relation of Reid (1991),
$M_V=2.89 + 3.37(V-I)$, with an error of $0.44$ mag.  

\section{Comments on Specific Fields}

	Several fields have specific features that require comment.
First, the Groth Strip ($l=96^\circ$, $b=60^\circ$) is a mosaic of 28
fields whose overall center is approximately given by the RA and Dec of row
4 in Table 1.
Because of a slight (7.5\%) overlap of neighboring fields, the total area is
slightly less than 28 isolated fields.  The magnitude limits are not precisely
uniform over this field.  There are sections each with 3.6\% of the total
area having detection thresholds that are 
0.21, 0.20, 0.16, 0.16, and $-1.16$ mag brighter than the modal value listed
in Table 1.  There are also sections each with 3.6\% of the total area having
saturation thresholds that are 0.05 and $-0.70$ mag brighter than the modal
value.  These differences are taken into account in the analysis.

	The actual saturation threshold of the Hubble Deep Field (HDF, 
$l=126^\circ$, $b=55^\circ$) is $I_\min=18.75$ 
(set by a minimum exposure of 1100 s).  However, unlike all of
the other fields in the sample, HDF was chosen as a ``blank field'' by first
inspecting Palomar Observatory Sky Survey plates (Williams et al 1996).  
Thus, it is 
possibly biased against ``bright'' stars that would be visible in these images.
We therefore choose a bright limit that is certainly fainter than the influence
of such possible bias, but is at the same time not influenced
by our knowledge of the stellar content of this particular field (as reported 
by Flynn et al.\
1996).  We choose the mode of the bright limits of the Paper I fields, 
$I_\min=19.45$, corresponding to a minimum exposure of 2100 s and to
$V_\min>21$ for the M dwarf sample selected by $V-I>1.53$.

	Three fields have less than full area coverage.  One chip (33\% of
the field) was unusable in the field at 
$l=140^\circ$, $b=26^\circ$.  Ten percent of the field at 
$l=77^\circ$, $b=36^\circ$ was also unusable.  One chip from the field at
$l=134^\circ$, $b=-65^\circ$ overlaps a neighboring field and is therefore
excluded.

\section{The Groth Strip}

\FIG\one{
Stars in the Groth Strip shown by $M_V$ (inferred from color) and height 
modulus above the plane ($\mu_z=V_0-M_V+5\log\sin b$).  Diagonal lines
represent the $I$-band range of sensitivity.  Large box is selection function
for M dwarfs described in text.  The number of stars predicted in each 
1 mag$^2$ box (such as the three small boxes at $M_V=12$) are the product
of the LF shown at the top (see Fig.\ 2) and the vertical density function
at right (see eq.\ 3.2 and Table 2).  Axis labels on the two smaller boxes
are short-hand and omit factors of $100$ and 
$2\times 10^{-3}(\ln 10)\Omega\csc^3 b$
respectively.  Small box at upper left contains
only spheroid subdwarfs. The actual number in the small box (19) is in good 
agreement with the model of Dahn et al.\ (1995), which predicts $20\pm 5$
stars in the box (see text).
}
\topinsert
\mongofigure{ps.fig1gx}{6.4}{6.7}{6.4}{
Stars in the Groth Strip shown by $M_V$ (inferred from color) and height 
modulus above the plane ($\mu_z=V_0-M_V+5\log\sin b$).  Diagonal lines
represent the $I$-band range of sensitivity.  Large box is selection function
for M dwarfs described in text.  The number of stars predicted in each 
1 mag$^2$ box (such as the three small boxes at $M_V=12$) are the product
of the LF shown at the top (see Fig.\ 2) and the vertical density function
at right (see eq.\ 3.2 and Table 2).  Axis labels on the two smaller boxes
are short-hand and omit factors of $100$ and 
$2\times 10^{-3}(\ln 10)\Omega\csc^3 b$
respectively.  Small box at upper left contains
only spheroid subdwarfs. The actual number in the small box (19) is in good 
agreement with the model of Dahn et al.\ (1995), which predicts $20\pm 5$
stars in the box (see text).
}
\endinsert

	Work on star counts in the modern era has proceeded by analyzing 
relatively
large fields each with many stars.  The resulting color-magnitude diagrams
can then be visually inspected and directly compared with results of the
models constructed from all the fields.  See Bahcall (1986) for a review.
By contrast, in the present program we analyze many small fields scattered
over the sky, most with very few stars.  Indeed, 47 of the 162 PC1 fields
contain no stars at all.  Hence visual comparison with models is not usually
fruitful.  Because of its large size, the Groth Strip offers a unique 
opportunity to compare the model with a field having many stars.  

	In Figure \one, we plot absolute magnitude (inferred from the observed
color: $M_V=2.89+3.37[V-I]$), and height modulus above the plane
$(\mu_z = V_0 - M_V + 5\log_{10}\sin b)$. Here $V_0=V$, since the extinction 
is $A_V=0$ (Burstein \& Heiles 1984).  The diagram can be compared
directly to figure 1 of Paper I.  The diagonal lines represent the magnitude
limits $I_\min$ and $I_\max$.  The large box is the M dwarf selection function
which is discussed in Paper I: $V-I>1.53$ (to avoid contamination by spheroid
giants), $V-I<4.63$ (to avoid the region where the color-mag relation becomes
double-valued), and $z<3200\,$pc (to avoid contamination by spheroid dwarfs).
The smaller rectangular figure at the top shows the LF 
(per 100 pc$^3$ per mag) taken from Figure 2, below.  The rectangular figure
at the right shows the
vertical distribution as given by equation (3.2) and Table 2, below.  The units
are chosen so that the expected density of stars in the diagram is the product
of the two numbers at the top and right.  For example, for the 3 small boxes
(each 1 mag$^2$), the LF is 1.0, and the height functions are 2.8, 4.2, and 5.0
respectively.  Hence, the  expected numbers of stars per box (or fraction of a
box where the box goes past the mag limits) are 2.8, 4.2, and 5.0, 
respectively.
With allowance for Poisson errors, the model is a good representation of the
data.

	The small box at the upper left of the main figure must contain 
only spheroid stars
because if these were disk stars they would be 6 to 8 kpc above the plane.
The observed kinetic energy of disk stars is insufficient to reach such
heights.
Using the spheroid LF of Dahn et al.\ (1995) and assuming that spheroid
stars in this color range are 2.5 mag fainter than disk stars of the same 
color (Monet et al.\ 1992), we predict that this box should contain a
total of $22 f$ stars, where $f$ is a number that depends on the spheroid
flattening, $c/a$.  For $c/a=0.6$, 0.8, and 1, we find $f=0.8$, 0.9, and 1.
This compares with 19 stars actually in the box.
We conclude that the Dahn et al.\ (1995) LF is consistent with the Groth
Strip data.  

	In Paper I, we argued that with the adopted height limit $z<3200$,
contamination by spheroid dwarfs is negligible and can be ignored.  We have
now checked that this is correct by incorporating into our models the spheroid 
stars expected on the basis of the Dahn et al.\ (1995) LF.  We find that indeed
the effects are much smaller than the Poisson errors, typically 
${\cal O}(1\%)$.  We therefore ignore spheroid contamination of the disk
star region.

\chapter{Results}

\section{Models}

	We model the distribution of stars as a function of Galactic position
and absolute magnitude by,
$$\Phi(i,z,R) = \Phi_i \nu(z) \exp\biggl(-{R-R_0\over H}\biggr),\eqn\phieq$$
where $\Phi_i$ is the LF at the Galactic plane for the $i$th magnitude bin,
$(R,z)$ is the Galactic 
position in cylindrical coordinates, $R_0=8\,$kpc is the
solar galactocentric distance, and $H$ is the disk scale length.  
We consider two forms for the vertical distribution function
$\nu(z)$, the ``$\rm sech^2$'' distribution
$$\nu_s(z)=(1-\beta)\sech^2{z\over h_1} + \beta\,
\exp{-|z|\over h_2},\eqn\nusech$$
and the ``double exponential'' distribution
$$\nu_e(z)=(1-\beta)\exp{-|z|\over h_1}+\beta\,\exp{-|z|\over h_2}.\eqn
\nuexp$$
The LF is assumed constant within each of
the nine luminosity bins centered at $M_V=8.25$
(1/2 mag), $M_V=9$, 10, 11, 12, 13, and 14 (1 mag), and $M_V=15.50$ and 17.50 
(2 mag).  (Later we also break up the last two bins into four 1-mag bins.)\ \ 
Thus, the fit has a total 13 parameters including nine LF parameters plus
$H$, $\beta$, $h_1$, and $h_2$.

\section{Global Parameters}

	We found in Paper I that the LF's for the $\sech^2$ and double
exponential fits were nearly identical up to an overall factor of 1.93 and
that for heights $z\gsim 300\,\pc$, $\Phi_{i,s}\nu_s(z)$ was nearly identical
to $\Phi_{i,e}\nu_e(z)$.  That is, the only difference between the two
distributions was the relative normalization near the plane.  Since the 
sample contains
almost no stars in this region (see fig.\ 1 from Paper I and also Fig.\ \one),
we could not distinguish between these models on the basis of the {\it HST}
data alone.  We therefore fixed the local normalization according to the LF
determined from local parallax stars (Reid, Hawley, \& Gizis 1995) in the
region $8.5\leq M_V\leq 12$ where all previous studies of the LF agree. We then
formed a linear combination of the two models to reproduce this density.
In fact, the $\sech^2$ model agreed almost perfectly with the local parallax
normalization and therefore was in effect the adopted model.  With the present
expanded data set we again find that the $\sech^2$ model agrees with the
parallax-star normalization to within 1\%.  We therefore simply adopt the
$\sech^2$ distribution
and dispense with linear interpolation.  

	Table 2 compares the best-fit parameters
of the model using all the available data with the previous values determined 
in Paper I.  The present and previous disk parameters are in excellent
agreement.  Here $\rho_0$ is  the mass density of M dwarfs at
the plane and $\Sigma_M$ is the column density of M dwarfs.

\midinsert
\singlespace
\overfullrule=0pt
\hoffset=0.375in
$$\vbox{\halign{#\hfil\quad&\hfil#\quad&\hfil#
\quad&\hfil#\hfil\quad&\hfil#\hfil\quad&\hfil#\hfil&\hfil#\hfil\cr
\multispan{7}{\hfil TABLE 2 \hfil}\cr
\noalign{\medskip}
\multispan{7}{\hfil Best-Fit Sech$^2$ Models for M Stars 
$(8<M_V<18.5)$\hfil}\cr
\noalign{\smallskip}
\noalign{\hrule}
\noalign{\smallskip}
\noalign{\hrule}
\noalign{\smallskip}
\hfil Data Set\hfil&\hfil$h_1$\hfil&\hfil$h_2$\hfil&
\hfil$\beta$\hfil&\hfil $\rho_0$\hfil
&\hfil$\Sigma_M$\hfil
&\hfil $H$\hfil\cr
&\hfil (pc)\hfil &\hfil (pc)\hfil && \hfil($M_\odot\rm pc^{-3}$)\hfil
 & \hfil($M_\odot\rm pc^{-2}$)\hfil
&\hfil (kpc) \hfil \cr
\noalign{\smallskip}
\noalign{\hrule}
\noalign{\smallskip}
Present  & $320\pm 50$ & $643 \pm 60$
& $21.6\pm 6.8$\% &$0.0158\pm 0.0041$ 
& $12.3\pm 1.8$
& $2.92\pm 0.40$ \cr
Previous  & $323\pm 54$ & $656 \pm 78$
& $19.8\pm 7.1$\% &$0.0159\pm 0.0044$ 
& $12.4\pm 1.9$
& $3.02\pm 0.43$ \cr
\noalign{\smallskip}
\noalign{\hrule}
}}
$$

\endinsert

\FIG\two{
M Dwarf LF as determined from {\it HST} star counts.  Triangles represent
present determination using maximum likelihood (ML) method including error
bars determined within the fit.  Circles represent the LF as determined
in Paper I by the same method.  Error bars not shown for these 
to avoid clutter.
As discussed in text, ML can amplify small scale structure caused by Poisson
fluctuations.  We therefore also show (crosses) the LF as determined by naive 
binning of star counts, a method which does not have this problem.  Error
bars for the naive-binning LF are shown only for the last four points.
}
\topinsert
\mongofigure{ps.fig2gx}{6.4}{6.0}{6.4}{
M Dwarf LF as determined from {\it HST} star counts.  Triangles represent
present determination using maximum likelihood (ML) method including error
bars determined within the fit.  Circles represent the LF as determined
in Paper I by the same method.  Error bars not shown for these 
to avoid clutter.
As discussed in text, ML can amplify small scale structure caused by Poisson
fluctuations.  We therefore also show (crosses) the LF as determined by naive 
binning of star counts, a method which does not have this problem.  Error
bars for the naive-binning LF are shown only for the last four points.
}
\endinsert

\section{Luminosity Function}

	It is not surprising that the global parameters change very little:
the total number of stars has increased by only 31\%.  Of greater interest
is the faint end of the LF ($13.5<M_V< 18.5$) whose determination depended
previously on only 23 stars and for which there are now 47.  Figure \two\
shows the present (triangles) and previous (circles) determinations of the LF.
The new data basically confirm the previous results.

	However, we have discovered a subtle interplay between statistical
and systematic effects which could affect the interpretation of the final
two points ($M_V=15.5$ and 17.5).  The maximum likelihood fitting procedure
(described in detail in Paper I) takes account of both measurement errors
and scatter of the absolute magnitudes (Malmquist bias).  If the
fitting procedure
finds that the counts in one bin are depressed relative to the bins on 
{\it both} sides, it will conclude that the intrinsic density in this bin
is {\it even lower} because more stars will have been scattered into than out
of the bin.  Thus, it will tend to further depress the LF at $M_V=15.5$
and raise it at $M_V=17.5$.  If the depression at $M_V=15.5$ is real (and not
simply a statistical fluctuation) and if the error estimates are accurate,
this will lead to a more accurate estimate of the LF at this bin.  However,
this bin contains a total of only 15 stars, and the final bin contains only
10.  Thus, the alternative explanation of a statistical fluctuation is
also plausible.  

	To investigate this possibility further, we make
the following alternative estimate of the LF.  We first calculate the
effective volume, $v_{\eff,j}(M_V)$, as a function of absolute magnitude
for each field $j$ by integrating the volume element as a function of
the distance $\ell$ along the line of sight,
$$v_{\eff,j}(M_V) = \Omega_j\int_{\ell_{-,j}(M_V)}^{\ell_{+,j}(M_V)} 
d \ell \,\ell^2
\nu_j[z(\ell)]\exp\biggl[{-{R(\ell)-R_0}\over H}\biggr].\eqn\veff$$
Here the integration limits $\ell_{\pm,j}(M_V)$ 
for the $j$th field are determined from the
sensitivity limits in Table 1, assuming that the color-mag relation holds
exactly with no intrinsic scatter and no errors.  The density function
$\nu(z)\exp[-(R-R_0)/H]$ is computed according to equation \nusech\ and
using the parameters given in Table 2.  Next, we form the total effective
volume from all fields, $v_\eff = \sum_j v_{\eff,j}$.
Finally, we divide the total number of stars in a given magnitude bin by
the effective volume integrated over that bin.  The results are shown as
crosses in Figure \two.

	We note that this method takes account of neither Malmquist bias
nor observational errors.  However, as we argued in Paper I, Malmquist bias
is not a significant problem for the {\it HST} survey because it extends
to the ``top'' of the disk.  In fact, one sees from Figure \two\ that this
simple binning procedure agrees quite well with the more sophisticated
fit over most of the LF.  We show the last four mags in individual 1-mag
bins.  We show (Poisson) error bars for these only to avoid clutter.
At the faint end, Poisson errors are a potentially much more serious problem
than Malmquist bias.  We see from Figure \two\ that the dip at $M_V=15.5$
could plausibly be a statistical deviation from an intrinsically smooth LF.
When transforming to a mass function, we therefore consider both the
maximum-likelihood and naive-binning determinations.

\FIG\three{
MFs derived from LFs 
on the basis of the mass-$M_V$ relation of Henry \& McCarthy
(1993).  Triangles and crosses show maximum-likelihood and naive-binning
fits as in Fig.\ \two.  Circles show MF based on the LF of Wielen et al.\
(1983) over the range $4\leq M_V\leq 11$.  The first 7 points for larger masses
are well fit by the indicated straight line,
$\Phi \equiv d N/d\log M = -1.37 - 1.21\,\log(M/M_\odot)$ for $M>0.6\,M_\odot$.
The {\it HST} crosses and the three relevant points from Wielen et al.\ (1983)
are well fit by the other straight line,
$\Phi = -0.99 + 0.44\,\log(M/M_\odot)$ for $M<0.6\,M_\odot$.  
Thus the power-law
index changes by 1.65 at $M\sim 0.6\,M_\odot$.
}
\topinsert
\mongofigure{ps.fig3gx}{6.4}{6.3}{6.4}{
MFs derived from LFs 
on the basis of the mass-$M_V$ relation of Henry \& McCarthy
(1993).  Triangles and crosses show maximum-likelihood and naive-binning
fits as in Fig.\ \two.  Circles show MF based on the LF of Wielen et al.\
(1983) over the range $4\leq M_V\leq 11$.  The first 7 points for larger masses
are well fit by the indicated straight line,
$\Phi \equiv d N/d\log M = -1.37 - 1.21\,\log(M/M_\odot)$ for $M>0.6\,M_\odot$.
The {\it HST} crosses and the three relevant points from Wielen et al.\ (1983)
are well fit by the other straight line,
$\Phi = -0.99 + 0.44\,\log(M/M_\odot)$ for $M<0.6\,M_\odot$.  
Thus the power-law
index changes by 1.65 at $M\sim 0.6\,M_\odot$.
}
\endinsert

\section{Mass Function}

	As in Paper I (see \S\ 4.1), we adopt the empirical mass-$M_V$ 
relation of Henry \& McCarthy (1993).  This relation is also in good agreement
with the theoretical results of Baraffe \& Chabrier (1996).  In Figure
\three\ we show the mass functions derived from the LF using maximum likelihood
(triangles) and naive binning (crosses).  Also shown  (circles)
is the MF derived from the LF of Wielen, Jahreiss, \& Kr\"uger (1983) on 
the basis of parallax stars in the range $4\leq M_V \leq 11$.

	The most important feature of Figure \three\ is the break in
the power law at $M\sim 0.6\,M_\odot$.  The straight line to the left is
a linear (i.e., power-law) fit to the first seven points from the 
Wielen et al.\ (1983) data.  The line to the right is a linear fit to the
{\it HST} data (crosses).  The respective slopes are $-1.21$ and $0.44$.
If the rising Salpeter-like slope to the left had remained unbroken, the four
lowest mass bins would be $\sim 20$ times more populated.
That is, they would contain $\sim 500$ stars rather than the 25 actually 
observed.

	The break in the power law is not an artifact of the
difference in surveys.  Note that the last three points of the Wielen et al.\
(1983) data which correspond to magnitude bins $M_V=9$, 10 , and 11, agree
well with the {\it HST} data.  This is true not only in the mean (which
is an effect of using the parallax-star LF in this region to determine the 
overall normalization) but also in their individual values.

	Note that the falling power-law is a reasonable fit to all of
the naively binned data points (crosses).  This suggests that the dip in 
the
LF at $M_V=15.5$ 
\break ($\log[M/M_\odot]\sim -0.1$) could plausibly be a statistical
fluctuation.  Nevertheless, the dip in the MF function is not purely an
artifact of the maximum likelihood method since it is reproduced (albeit
with reduced intensity) by the naively binned data.  The sample is 
not large enough to determine if this dip is real.

\section{Binary Correction}

	The {\it HST} data are sensitive to binaries with separations
$\gsim 0.\hskip-2pt''3$ which, at typical distances of $\sim 2\,$kpc, 
corresponds to projected separations $\gsim 600\,$AU.  Since only 1--2\%
of stars have binary companions in this range (Gould et al.\ 1995), while
half or more of all stars are in binaries, the {\it HST} sample misses 
essentially all secondaries in binary systems.

	We can make an empirical estimate of the incompleteness due
to missed secondaries using the results of Reid et al.\ (1995).  Reid et al.\
 divided
nearby M stars into two groups: (1) primaries and isolated stars, and (2)
secondaries.  (See also fig.\ 2 of Paper I).  The nearby sample should be 
relatively complete because the stars are studied both photometrically and 
spectroscopically.  For early M dwarfs ($M_V\leq 11$, $\log(M/M_\odot)> -0.4$),
about 10\% of stars are secondaries and so would be missed by {\it HST}.
For late M dwarfs ($M_V\geq 15$, $\log(M/M_\odot)< -0.9$), about half are
secondaries.  Thus, incorporating the missing secondaries should change the
slope from 0.44 to $\sim 0.1$.  While this correction is admittedly crude,
it is clear that missing binary companions cannot account for the 
break in power law.  

\section{Disk Mass}

	Our estimate for the total column density of M stars remains
unchanged from Paper I (see Table 2) and therefore so does our estimate
of the total column density of the disk: 
($\Sigma\simeq 40\,M_\odot\,\pc^{-2}$).
As we noted there, this is significantly lower than all published estimates of
the dynamical mass of the disk.  If one were to assume that the observed mass
function (see Fig.\ \three) continues into the brown dwarf regime with slope 
$\alpha$ and normalized by twice the value of the last point (to take account
of binaries), then the total column density of brown dwarfs would be 
$\Sigma_{\rm BD} = 6 M_\odot/(1+\alpha)$.  For $\alpha\sim 0$, this would 
bring the disk column density into line with the lowest dynamical estimate
(Kuijken \& Gilmore 1989), but not with most others (e.g.\ Bahcall 1984;
Bienaym\'e, Robin, 
\& Cr\'ez\'e 1987; Bahcall, Flynn, \& Gould 1992; Flynn \& Fuchs 1994).
To reach the highest estmate of $\sim 80\,M_\odot\,\pc^{-2}$ 
(Bahcall et al.\ 1992) would require
a dark component of $\sim 40\,M_\odot\,\pc^{-2}$ which would be produced,
for example, by $\alpha=-2$ and a cutoff at $M=0.01\,M_\odot$.

\section{Microlensing}

	The optical depth to microlensing toward the Large Magellanic Cloud 
(LMC, $b=-33^\circ$) of a general disk distribution $\rho(z)$ is given
by $\tau = \csc^2 b\int_0^\infty d z\,4 \pi G\rho(z)z/c^2$.  For a $\sech^2$ 
and exponential distributions, this becomes 
$\tau = 2(\ln 2)\pi G\Sigma h\csc^2 b/c^2$, and 
$\tau = 2\pi G\Sigma h\csc^2 b/c^2$, respectively. 
Using these formulae and the parameter
values given in Table 2, and assuming that the total column density of stars 
is $\Sigma_{\rm tot}=2.1\Sigma_M$ (see Paper I), we estimate that the 
optical depth due to disk stars is 
$\tau\sim 8\times 10^{-9}$, a factor
of 25 lower than the most recent estimate for the observed optical depth
$\tau\sim 2\times 10^{-7}$ by the MACHO collaboration (Alcock et al.\ 1996).  

	We note that the MACHO collaboration's very high magnification event 
(number 5 in Alcock et al.\ 1996) 
could well be due to a disk star.  The color and magnitude of the unlensed
light in this event are inconsistent with the characteristics of any known
population in the LMC, but are consistent with a foreground late M dwarf
within the seeing disk of the lensed source (an LMC main-sequence star). 
The probability of finding an M dwarf in a randomly chosen seeing disk is 
small, but an M dwarf would be the most likely stellar type recovered
in this way if the lens were a disk foreground star.

	The break in the slope of the MF shown in Figure \three\ has important
implications for the interpretation of microlensing toward the Galactic
Bulge.  Zhao, Spergel, \& Rich (1995) have shown that the bulge event rate
would be well explained if all of the dynamically-measured bulge mass 
($\sim 2\times 10^{10}M_\odot$) were in a Salpeter MF ($\alpha=-1.35$) over 
the range $0.08 M_\odot\leq M\leq 0.6 M_\odot$.  That is, the events could be 
explained without recourse to brown dwarfs or other dark matter.  The bulge 
LF has now been measured down to $V=26$ ($M_V\sim 10$ or $M\sim 0.5\,M_\odot$) 
by Light, Baum, \& Holtzman (1996) and these observed stars account for
$\sim 1 \times 10^{10}M_\odot$ of the bulge mass but very few of the lensing
events (Han 1996).  The shape of the observed bulge LF is consistent with that
of the
disk LF over the same range and is therefore consistent with a Salpeter MF.  
Han \& Gould (1995) showed that a Salpeter 
MF explains the bulge events provided
that it is extended to $\sim 0.05\,M_\odot$, i.e., includes a substantial
number of brown dwarfs.  The low-mass end of the disk MF is, however, 
inconsistent with a Salpeter MF (see Fig.\ \three).  While the
agreement between disk and bulge LFs in the region where they are both
observed is no guarantee of their agreement at fainter mags, similar LFs
(and therefore MFs) does appear to be the simplest hypothesis.  If the
bulge MF is extended using the estimate for the disk MF given in Paper I,
then the observed+inferred bulge stellar population accounts for $1.4\times
10^{10}M_\odot$, i.e.\ $\sim 70\%$ of the dynamical estimate, but these stars
still account for only a small fraction of the observed microlensing events.
Han (1996) finds that the remaining $\sim 30\%$ must be in brown dwarfs to
explain the observed events.  In brief, if current estimates of the lensing
rate toward the bulge are confirmed then either: 1) the bulge and disk mass
functions are very different for $0.1\,M_\odot \leq M \leq 0.5\,M_\odot$
or 2) the bulge contains a large population of brown dwarfs.

\chapter{Future Prospects}

	There are two types of additional observations that would significantly
improve our understanding of disk M dwarfs.
The first is to acquire more statistics, 
particularly for late M dwarfs.  We have demonstrated the existence of a 
break in the power law of the disk MF near $M\sim 0.6\,M_\odot$.  
However, more data are required to investigate the detailed structure of the
MF.  A large body of suitable {\it HST} WFC2 observations have
already been made and we are in process of analyzing these.  More observations
are expected in the future.

	Second, the Groth Strip offers a unique opportunity for ground-based
follow-up observations.  In most cases, it is not efficient to observe 
individual deep WFC2
fields from the ground because of the ${\cal O}(10^2)$ mis-match in field 
sizes.  One consequence of this mis-match
is that there are no empirical calibrations
of the {\it HST} filters for late M dwarfs.  The Groth Strip contains over
50 disk M dwarfs (plus an equal number of spheroid M dwarfs) and would 
therefore provide an excellent empirical check on the calibration of Bahcall 
et al.\ (1994) which was determined by convolving the {\it HST} filter
functions with ground-based spectra.  In addition, deep $B V I$ observations
would allow one to measure the $B$ excess (at fixed $V-I$) of each star
in the field and so estimate the degree to which any of the stars
may be sub-luminous.
This would permit more sophisticated modeling of the luminosity of stars
than the single color-mag relation used here.  At present, there are no
data to form the basis for such modeling and theoretical estimates are too
uncertain to be of use.  It would be possible to make such observations in
non-{\it HST} fields and P.\ Boeshaar (private communication, 1996) is 
carrying out a study of this type.
  However, observation of the Groth Strip has the important
advantage that the stars are already identified unambiguously.  In addition,
as mentioned above, these observations would be very valuable for calibration
of the {\it HST} filters.

{\bf Acknowledgements}:  A.\ G.\ was supported in part
by NSF grant AST 9420746 and in part by NASA grant NAG5-3111.
J.\ N.\ B.\ was supported by NASA grant NAG5-1618.
The work is based in large part
on observations with the NASA/ESA Hubble Space Telescope, obtained
at the Space Telescope Science Institute, which is operated by the
Association of Universities for Research in Astronomy, Inc. (AURA), under
NASA contract NAS5-26555.  Important supplementary observations were made
at KPNO and CTIO operated by AURA.

\endpage

\Ref\Bdmo{Bahcall, J.\ N.\ 1984, ApJ, 276, 169}
\Ref\Bst{Bahcall, J.\ N.\ 1986, ARA\&A, 24, 577}
\Ref\BFG{Bahcall, J.\ N., Flynn, C., \& Gould, A.\ 1992, ApJ, 389, 234.}
\Ref\BFG{Bahcall, J.\ N., Flynn, C., Gould, A., \& Kirhakos, S.\ 1994, ApJ, 
435, L51.}
\Ref\BC{Baraffe, I., \& Chabrier, G.\ 1996, ApJ, 461, L51}
\Ref\BRC{Bienaym\'e, O., Robin, A., \& Cr\'ez\'e, M.\ 1987, A\&A, 180, 94}
\Ref\bh{Burstein, D.\ \& Heiles, C.\ 1984, AJ, 90, 817}
\Ref\Cow{Cowley, A.\ P., \& Hartwick, F.\ D.\ A.\ 1991, ApJ, 373, 80}
\Ref\dah{Dahn, C.\ C., Liebert, J.\ W., \& Harrington R. S.\ 1986, AJ, 91, 621}
\Ref\dah{Dahn, C.~C., Liebert, J.\ W., Harris, H., \& Guetter, H.\ C.\ 1995, 
p.\ 239, An ESO Workshop on: the Bottom of the Main Sequence and Beyond,
C. G. Tinney ed. (Heidelberg: Springer)}
\Ref\ff{Flynn, C.\ \& Fuchs, B.\ 1994, MNRAS, 270, 471}
\Ref\ff{Flynn, C.\ \& Gould, A., \& Bahcall, J.\ N., 1996, ApJ, 466, L55}
\Ref\gbm{Gould, A., Bahcall, J.\ N., \& Flynn, C.\ 1996, ApJ, 465, 759}
\Ref\gbm{Gould, A., Bahcall, J.\ N., Maoz, D., \& Yanny, B.\ 1995, ApJ, 441, 
200}
\Ref\han{Han, C.\ 1996, ApJ, submitted}
\Ref\han{Han, C.\ \& Gould, A.\ 1996, ApJ, 467, 540}
\Ref\hmc{Henry, T.\ J.\ \& McCarthy, D.\ W.\ 1993, AJ, 106, 773}
\Ref\KG{Kuijken, K.\ \& Gilmore, G. 1989, MNRAS, 239, 605}
\Ref\light{Light, R.\ M., Baum, W.\ A., \& Holtzman, J.\ A.\ 1996, 
in preparation}
\Ref\mon{Monet, D.\ G., Dahn, C.\ C., Vrba, F.\ J., Harris, H.\ C.,
Pier, J.\ R., Luginbuhl, C.\ B., \& Ables, H.\ D.\ 1992, AJ, 103, 638}
\Ref\reid{Reid, I.\ N.\ 1991, AJ, 102, 1428}
\Ref\rei{Reid, I.\ N.,  Hawley, S.\ L., \& Gizis, J.\ E.\ 1995, AJ, 110, 1838}
\Ref\will{Williams, R.\ E., et al.\ 1996, AJ, 112, 1335}
\Ref\WJK{Wielen, R., Jahreiss, H., \& Kr\"uger, R.\ 1983, IAU Coll.\ 76:
Nearby Stars and
the Stellar Luminosity Function, A.\ G.\ D.\ Philip and A.\ R.\ Upgren eds.,
p 163}
\refout
\endpage
\end
\topinsert
\special{ps: plotfile figure1.post}
\endinsert
\topinsert
\special{ps: plotfile figure2.post}
\endinsert
\bye